\documentclass[%
 amsmath,amssymb,
 reprint
]{revtex4-1}
 
\usepackage{graphicx}
\usepackage{dcolumn}
\usepackage{bm}

\usepackage[utf8]{inputenc}
\usepackage[T1]{fontenc}
\usepackage{mathptmx}
\usepackage{etoolbox}
\usepackage{amsmath}

\makeatletter
\def\@email#1#2{%
 \endgroup
 \patchcmd{\titleblock@produce}
  {\frontmatter@RRAPformat}
  {\frontmatter@RRAPformat{\produce@RRAP{*#1\href{mailto:#2}{#2}}}\frontmatter@RRAPformat}
  {}{}
}%
\makeatother
\begin{document}

\title{Quantum non-Markovianity, quantum coherence and extractable work in a general quantum process}

\author{Amin Mohammadi}
\author{Afshin Shafiee*}
 \email{shafiee@sharif.edu}

\affiliation{ 
Research Group on Foundations of Quantum Theory and Information, Department of Chemistry,
Sharif University of Technology P.O.Box 11365-9516, Tehran, Iran}
\begin{abstract}
A key concept in quantum thermodynamics is extractable work, which specifies the maximum amount of work that can be extracted from a quantum system. 
Different quantities are used to measure extractable work, the most prevalent of which are ergotropy and the difference between the non-equilibrium and equilibrium quantum free energy. Using the latter, we investigate the evolution of extractable work when an open quantum system goes through a general quantum process described by a completely-positive and trace-preserving dynamical map. We derive a fundamental equation of thermodynamics for such processes as a relation between the distinct sorts of energy change in such a way the first and the second law of thermodynamics are combined. We then identify the contributions from the reversible and irreversible processes in this equation and demonstrate that they are respectively responsible for the evolution of heat and extractable work of the open quantum system. Furthermore, we show how this correspondence between irreversibility and extractable work has the potential to provide a clear explanation of how the quantum properties of a system affect its extractable work evolution. Specifically, we establish this by directly connecting the change in extractable work with the change in standard quantifiers of quantum non-Markovianity and quantum coherence during a general quantum process. We illustrate these results with two examples. 
\end{abstract}

\maketitle

\section{\label{sec:level1}Introduction}

Thermodynamics can be thought of as an extension of classical mechanics to contain concepts such as temperature and heat in the study of properties of macroscopic systems \cite{lebon2008understanding}.
Depending on the underlying mechanics,
one may wonder how the transition from classical to quantum mechanics will affect the laws of thermodynamics. It becomes even more intriguing if the system under study exhibits quantum effects with no classical counterpart. Followed by advances in laboratory and quantum technological implementations, much attention in recent years has been devoted to understanding thermodynamics of small quantum systems \cite{gemmer2009quantum,vinjanampathy2016quantum,binder2018thermodynamics,goold2016role,deffner2019quantum,horodecki2013fundamental,brandao2015second,cwiklinski2015limitations,linden2010small,scully2003extracting,masanes2017general,taranto2023landauer,kammerlander2016coherence,perarnau2015extractable,taranto2023landauer,alipour2022entropy,soltanmanesh2020can,soltanmanesh2019clausius,myers2022quantum,bhattacharjee2021quantum,quan2007quantum,dann2020quantum}.
One of the most fundamental challenges in this new field of research is finding the general definitions of work and heat for arbitrary quantum systems.
Although standard formulations of work and heat exist with limitations to weak coupling conditions and Markovian dynamics \cite{alicki1979quantum}, the situation in general quantum evolutions is somewhat unclear.
The difficulty arises from the fact that both work and heat are path-dependent quantities that can not be described by Hermitian operators \cite{talkner2007fluctuation},
a standard way to define physical observables in quantum mechanics.    
In the absence of a single, universal concept, several definitions based on various criteria emerge.
For example, a variety of approaches from use of quantifiers like ergotropy \cite{allahverdyan2004maximal} and free energy difference \cite{skrzypczyk2014work,muller2018correlating,garcia2020fluctuations} for work extraction from non-passivity of quantum states to statistical treatments attempting to measure quantum work during a process by constructing appropriate probability distributions \cite{esposito2009nonequilibrium,seifert2012stochastic,campisi2011colloquium} have been proposed.

In this regard,
in a remarkable study Binder \emph{et al.} \cite{binder2015quantum} established in an operational sense a relation in resemblance to the first law of thermodynamics between work done, extractable work, heat and change of internal energy for a quantum system evolving in a general quantum process presented by a completely-positive and trace-preserving (CPTP) map.  
The constituents of this relation are the difference in ergotropy between the initial and final states of the map, adiabatic work and introduced operational heat that their sum is equal to the change in internal energy of the open system due to general quantum process.

In the present study,
we establish similarly an energy balance relation for general quantum processes in which the free energy work is used to measure the extractable work from the system during the process. This relation replaces the abovementioned operational heat with the change in entropy of the system due to the fact that in contrast to ergotropy, the free energy work exploits the resource from open quantum systems.
We interpret this relation as the fundamental thermodynamic equation combining the first and the second law of thermodynamics for general quantum processes.
We then show
the time evolution of heat and the extractable work are determined
respectively by the reversible and irreversible change in the
entropy of the system that allows us to obtain, based on pure thermodynamic grounds, the unambiguous  definitions of heat and work in the general quantum processes.
The fact that irreversibility governs the evolution of the extractable work underlies the main results of this paper.
We demonstrate how this fact can be used to link directly quantum non-Markovianity in time evolution of the open system with the rate at which its extractable work changes.
In addition,
we discuss how such a utility is also provided for exploring the connection between the amount of extractable work and the time evolution of quantum coherence.
Therefore, our results clearly demonstrate the resource nature of quantum non-Markovianity and quantum coherence in quantum thermodynamics.
In the end,
as an illustration of our results, we investigate the effects of quantum non-Markovianity and quantum coherence on the evolution of the charging power for two types of quantum batteries: a two-qubit battery-charger model and a two-qubit battery charged by quantum coherence due to one photon mode mediation.

In the remainder of the paper, we first briefly review the concept of ergtropy and what is called the operational first law of quantum thermodynamics in the subsequent section before presenting our results in sections III and IV. Examples are discussed in section V and the paper is ended with the conclusions in section VI. 

\section{Ergotropy and first law of quantum thermodynamics}
The first Law of Thermodynamics shows the contribution of work and heat to internal energy change in a thermodynamic process.
In quantum thermodynamics, depending on whether one considers a system out of passive or thermal equilibrium (completely passive) states as a resource from which work can be extracted,
ergotropy and free energy difference are the two most widely used extractable work quantifiers.    
More precisely,
ergotropy \cite{allahverdyan2004maximal} quantifies the maximum work that can be extracted from a system by transforming it from a non-passive to a passive state in a cyclic unitary manner
\begin{equation}
W_e = Tr[\rho \emph{H}-\pi \emph{H} ]
\label{eq:1}.
\end{equation}
The passive state $\pi$ is represented as $\pi= \Sigma$ $r_n |\varepsilon_n><\varepsilon_n|$ provided that the state $\rho$ and Hamiltonian \emph{H} are expressed in their spectral decomposition respectively in a decreasing and increasing order i.e., $\rho=\Sigma r_n |r_n><r_n|$ and $\emph{H}=\Sigma \varepsilon_n |\varepsilon_n><\varepsilon_n|$ with $r_{n+1}\leq r_n $ and $\varepsilon_{n+1}\geq \varepsilon_n$ $\forall n$.    
Cyclicity,
here,
refers to the fact that Hamiltonian is identical in the initial and final points of evolution. The ergotropy has been measured experimentally in the quantum heat engines with spin \cite{van2020single} and single atom \cite{von2019spin} as working fluids and recently in quantum batteries modeled by low dimensional metal complexes \cite{cruz2022quantum}.
An extension of ergotropy for open quantum systems and non-unitary evolutions has been made by Binder \emph{et al.} \cite{binder2015quantum}.
They recovered the first law of thermodynamics for finite quantum systems undergoing general quantum processes: 
\begin{equation}
\Delta{E} = \Delta{W_e}+<W>_{ad}+<Q>_{op} 
\label{eq:2}.
\end{equation}

Here, $\Delta{W_e}=W_e$($\emph{H}_\tau,\rho_\tau$) - $W_e$($\emph{H}_0$,$\rho_0$) is change in ergotropy of the system during the process ($\emph{H}_0$,$\rho_0$)$\to$($\emph{H}_\tau,\rho_\tau$) 
in which
$\rho_0$ and $\rho_\tau$ 
are initial and final quantum states that denote inputs and outputs of CPTP map $\Lambda_\tau$ i.e. $\rho_\tau =\Lambda_\tau \rho_0$ 
and
$\emph{H}_0$ and $\emph{H}_\tau$  
are initial and final Hamiltonians
with corresponding passive states $\pi_\tau$ and $\pi_0$.
Although
 $\Delta{W_e}$ bears much resemblance to the change of thermodynamic state functions as it depends only on the states and Hamiltonians in the initial and final of a process, here, it enters a genuine out-of-equilibrium contribution in the first law relation. This is important because any generalization of thermodynamic laws to quantum regimes must have the ability to describe both equilibrium and non-equilibrium situations.

 Also, $<W>_{ad}$ measures the change in internal energy due to a non-cyclic unitary transformation that change in Hamiltonian is assumed to be adiabatic in a quantum sense, i.e. at each instant of evolution eigenstates of Hamiltonian remain eigenstates 
 \begin{equation}
 <W>_{ad}= Tr[\pi_t \emph{H}_\tau]-Tr[\pi_m \emph{H}_0]
 \label{eq:3}
\end{equation}
where $\pi_\tau$ and $\pi_m$ are passive states with respect to final and initial Hamiltonians.
The subscript m is used to distinguish between states of $\pi_0$ and $\pi_m$ as they are passive states with different eigenvalues $\{r_n\}$ in the eigenbasis of initial Hamiltonian $\emph{H}_0$.
 Then, any energy change in the transformation between two passive states $\pi_0$ and $\pi_m$ 
when the Hamiltonian $\emph{H}_0$ remains unchanged can be interpreted as an operational heat (for example an equilibrium heating, see \cite{binder2015quantum}) accounting for the heat contribution in the first law formulation of quantum thermodynamics as indicated by the last term in Eq.~(\ref{eq:2}): 
\begin{equation}
<Q>_{op}= Tr[\pi_m \emph{H}_0]-Tr[\pi_0 \emph{H}_0].
\end{equation}
. 

\section{Free energy and fundamental equation of quantum thermodynamics}
Free energy is one of the most important quantities in thermodynamics because, according to Clausius's statement of second law, it informs us of the possible state transitions of a system and the optimal work can be extracted from the system as a result. In consequence of generally extending the validity of second law (in Clausius form) to quantum regime \cite{vinjanampathy2016quantum}, one can consider another extractable work quantifier that is given by the free energy difference between non-equilibrium and equilibrium states of the system in contact with a thermal bath \cite{skrzypczyk2014work,muller2018correlating,garcia2020fluctuations}
\begin{equation}
{W_f} = \emph{F}(\rho)-\emph{F}(\pi^\beta)
\end{equation}
where  $\pi^\beta = exp(-\beta \emph{H})/Z$ is the thermal equilibrium or Gibbs state with inverse temperature $\beta$ and partition function Z and $\emph{F}(\rho)=tr[\emph{H}\rho]-\beta^{-1} \emph{S}(\rho)$ is free energy with $S$ denoting the entropy and $\emph{F}(\pi^\beta)$ is equilibrium free energy of the system. 
There have been some protocols to extract this kind of quantum work in both individual \cite{skrzypczyk2014work,muller2018correlating} and asymptotically many copies of quantum systems \cite{brandao2013resource}.   
Considering the general process $(\emph{H}_0,\rho_0)\to (\emph{H}_\tau,\rho_\tau =\Lambda_\tau\rho_0)$, we write the change in the free energy work in initial and final points of evolution:  
\begin{equation}
\Delta{W_f} = {W_f}_\tau - {W_f}_0
\end{equation}
from which we obtain the following relation for a general time-dependent Hamiltonian:
\begin{equation}
\Delta{E} = \Delta{W_f}+<W>_{ad}+\beta^{-1} (\Delta{S}_(\rho)-\Delta{S}(\pi^\beta))
\label{eq:7}.
\end{equation}
As Eq.~(\ref{eq:2}) was described as the first law of thermodynamics for general quantum processes,  we interpret Eq.~(\ref{eq:7}) as the fundamental thermodynamic equation for such processes that adds to the power of thermodynamic description with combining the first and the second law.
The first and second terms similar to Eq.~(\ref{eq:2}) are extractable and adiabatic works with the difference that the former is quantified by free energy difference instead of ergotropy and in the latter the instantaneous Gibbs equilibrium states is singled out among possible passive states of system at the initial and final times of transformation. The main difference, however, is the third term that shows the change in energy which alters von Neumann entropies of the system :
\begin{equation}
\Delta{S}(\rho)= S(\rho_\tau)-S(\rho_0)
\label{eq:8}.
\end{equation}
\begin{equation}
\Delta{S}(\pi^\beta)= S(\pi_\tau^\beta)-S(\pi_0^\beta)
\label{eq:9}.
\end{equation}
We note that the above equations cannot define the heat in a general CPTP evolution before assigning the reversible and irreversible contributions in entropy change. To this end, we add and subtract the expression $Tr(\rho_0 log(\pi^\beta_0) +\rho_\tau log(\pi^\beta_\tau))$ to Eqs.~(\ref{eq:8}) and (\ref{eq:9}) through that we can write $\Delta{S}(\rho)-\Delta{S}(\pi^\beta)=\Delta{S}_{Ir}-\Delta{S}_R$ by the following definitions:
\begin{equation}
 \Delta{S}_{Ir}= S(\rho_0||\pi^\beta_0)- S(\rho_t||\pi^\beta_\tau)
 \label{eq:10}
\end{equation}
and
\begin{equation}
\Delta{S}_R = Tr[(\rho_\tau-\pi^\beta_\tau)log(\pi^\beta_\tau)]-Tr[(\rho_0-\pi^\beta_0)log(\pi^\beta_0)]
\label{eq:11}.
\end{equation}

We have defined the change in irreversible entropy in Eq.~(\ref{eq:10}) as the change in the initial and final values of quantum relative entropy ($S(\rho||\pi)= Tr[\rho (log(\rho)-log(\pi))]$) between non-equilibrium and corresponding equilibrium states of open system during the general quantum process.
Quantum relative entropy is a useful function in the quantum information theory because its output provides a measure of the distance between its two input states \cite{nielsen2010quantum}.
Therefore, Eq.~(\ref{eq:10}) characterizes the irreversible nature of the quantum dynamics of open system by expressing how thermodynamic equilibrium is approached for open system throughout the quantum dynamics. 
In the case that the Hamiltonian of the system remains unchanged during the process, Eq.~(\ref{eq:10}) turns into $\Delta{S}_{Ir}= S(\rho_0||\pi^\beta)- S(\rho_\tau||\pi^\beta)$, with $\pi^\beta$ being the unique equilibrium state of the system,   
that is a famous expression for entropy production indicating the irreversible term in the entropy balance relation.
Because of the relative entropy's contractivity property when subject to the action of CPTP maps \cite{vedral2002role}, the irreversible entropy change defined by Eq.~(\ref{eq:10}) is a positive amount. It may, however, be negative for time-independent Hamiltonians, as was the case with CPTP non-Markovian dynamics with a non-stationary thermal equilibrium state \cite{marcantoni2017entropy}.      
On the other hand, we have introduced Eq.~(\ref{eq:11}) which includes the change in the state of the system between non-equilibrium and equilibrium situations as the reversible change in entropy. 
It may also be viewed as an extension of the quantum Hatano-Sasa inequality \cite{hatano2001steady,vinjanampathy2015entropy}, a well-known formulation of the second law for CPTP maps which is expressed by $\Delta{S}\geq Tr[(\rho_\tau-\rho_0)log(\pi^\beta)]$, to CPTP maps with time-dependent Hamiltonians. In other words, Eqs.~(\ref{eq:10}) and (\ref{eq:11}) extend the existing definitions of $\Delta{S}_{Ir}$ and $\Delta{S}_{R}$ to the more general scenario where the system does not reach a unique equilibrium state in general due to the change in the Hamiltonian over time interval [0, $\tau$], but the instantaneous equilibrium states can be taken into consideration.

Putting the expressions for $\Delta{S}_R$ and $\Delta{S}_{Ir}$ (Eqs.~(\ref{eq:10}) and (\ref{eq:11})) into Eq.~(\ref{eq:7}), we obtain the following definition for heat flow from the system corresponding to the reversible change in entropy
\begin{equation}
 <Q> = -\beta^{-1}\Delta{S}_{R}= \Delta{E}-<W>_{ad}
 \label{eq:12}
\end{equation}
and moreover, find the time evolution of extractable work for the system is governed by irreversible entropy change, that is:  
\begin{equation}
\Delta{W_f}= -\beta^{-1}\Delta{S}_{Ir}.
\label{eq:13}
\end{equation}

We emphasize the importance of our final results in this section (Eqs.~(\ref{eq:12}) and (\ref{eq:13}))  as they show that for any physically legitimate transformation between states of an open quantum system, there are general, unambiguous definitions for heat and work in terms of well-defined thermodynamic quantities and processes. Furthermore, Eq.~(\ref{eq:13}) allows us to state our main result in the next section: providing a clear picture of the resource character of the quantum non-Markovinity and quantum coherence in quantum thermodynamics.    

\section{Effective Quantum Parameters on Extractable work of open quantum systems}
We begin this section by noting the rate at which the extractable work changes can be defined concerning Eq.~(\ref{eq:13}) as:
\begin{equation}
    P(t)=\lim_{\Delta{t} \to 0} \frac{\Delta{W_f}}{\Delta{t}} = -\beta^{-1}\frac{d{S}_{Ir}}{dt}
    \label{eq:14}.
\end{equation}
$P(t)$ is commonly referred to as charging power in the study of quantum batteries, which employ $P(t)$ to investigate how quickly a battery may be charged or discharged \cite{garcia2020fluctuations}. 
We then go over how Eqs.~(\ref{eq:13}) and (\ref{eq:14}) can be used to highlight the thermodynamic importance of particular quantum properties such as non-Markovianity and coherence as valuable resources on account of the enhancement of extracted work.

\subsection{\label{sec:level2}Quantum non-Markovianity}

Irreversibility is a fundamental concept in the theory of open quantum systems  \cite{breuer2002theory}. 
The interaction between the system and the environment causes the system to continuously lose its quantum properties, which is typically understood as an irreversible flow of information from the system to the environment. However, in some situations, for instance, when there is a strong system-environment interaction, the information that the system loses to the environment can be recovered. In contrast to Markovian systems, which lose information irreversibly, these systems that revive information are referred to as non-Markovian open quantum systems \cite{breuer2002theory}.
Several approaches have been proposed in the literature to quantify non-Markovian character of the open quantum system dynamics mainly based on the mathematical properties of the dynamical map \cite{rivas2010entanglement,chruscinski2014degree}  or physical features of the system-environment interaction \cite{breuer2009measure,liu2015quantifying,luo2012quantifying} (see also \cite{breuer2016colloquium,rivas2014quantum} for two reviews).
Since quantum properties are well known to be valuable work resources, it is anticipated that reviving these properties and resulting in non-Markovianity will have an advantage in thermodynamic studies. For example, some studies report positive effects of non-Markovianity on quantum Landauer erasure work cost revivals \cite{bylicka2016thermodynamic}, the performance of various kinds of quantum heat engines \cite{zhang2014quantum,abiuso2019non,camati2020employing,ptaszynski2022non}, and the charging of quantum batteries \cite{kamin2020non,ghosh2021fast,tabesh2020environment}.

On the other hand, according to thermodynamics,
irreversibility is considered as a contribution to the entropy balance relation of the system  \cite{lebon2008understanding,callen1991thermodynamics}. As we can see from Eq.~(\ref{eq:13}), this contribution determines the change in extractable work from the system. The extractable work calculated from this relation is monotonically decreasing throughout a CTPT evolution as a result of an increase in irreversibility, which is indicated by a positive value for $\Delta{S}_{Ir}$.
However, the the rate of change i.e. charging power can become in general negative in some time intervals.

In this respect, we note a precise connection with non-Markovianity is deducible when we take charging power into account and apply the result of Chen \emph{et al.} \cite{chen2017thermodynamic} to measure the amount of non-Markovianity during evolution. 
Based on the reconciling the concept of irreversibility in the theory of open quantum systems and quantum thermodynamics, Chen \emph{et al.} show that under the hypothesis of static environment, the non-Markovianity of open quantum systems can be defined as:
\begin{equation}
    \emph{I}=-\frac{\partial S_{Ir}} {\partial t}
    \label{eq:15}.
\end{equation}
The static hypothesis is satisfied if the environment state does not change during evolution. Now the charging power can be simply written as:
\begin{equation}
    P(t)=\beta^{-1} \emph{I}
    \label{eq:16}.
\end{equation}
 This result demonstrates that, in the absence of additional quantum resources, non-Markovianity is a necessary resource to achieve more thermodynamic efficiency in the quantum regime. 
 In other words, Eq.~(\ref{eq:16}) evidently shows that the more quantum properties a system revives, as indicated by a higher positive amount of non-Markovianity measure $\emph{I}$, the more charging power it recovers. The non-Markovianity advantage is especially apparent when the system is thermalized using baths at a higher temperature.

 \subsection{\label{sec:level2}Quantum coherence}

 Quantum coherence,
both fundamentally and practically,
is one of the most important quantum features as it distinguishes the quantum world from the classical one and is an essential resource in many quantum processes \cite{streltsov2017colloquium}.
When defined with relative to energy basis, quantum coherence is a prominent resource for carrying out certain thermodynamic tasks. Many studies have been conducted in recent years to look into the role that coherence play in the work extraction protocols \cite{lostaglio2015description,lostaglio2015quantum,korzekwa2016extraction,lostaglio2019introductory}, operation of quantum thermal machines \cite{rahav2012heat,scully2011quantum,mitchison2015coherence,uzdin2015equivalence}, the performance of quantum batteries\cite{garcia2020fluctuations,kamin2020entanglement,mayo2022collective,shi2022entanglement}, etc.
Here, we show that Eq.~(\ref{eq:13}) is capable of efficiently interpreting the impact of a system's quantum coherence on the work that can be extracted from it.  
This is provided by a direct connection between the change in free energy work and relative entropy of coherence, a widely used measure of quantum coherence \cite{baumgratz2014quantifying} defined as:
\begin{equation}
    C_{r}(\rho)= S(\Delta \rho)-S(\rho)
    \label{eq:17}
\end{equation}
where $\Delta\rho$ denotes the state with the same diagonal elements as $\rho$ but zero off-diagonal ones.
The relative entropy of coherence was first introduced in the context of resource theory of coherence \cite{baumgratz2014quantifying}, to quantify the resourcefulness of a coherent state and satisfies a set of required conditions, prime among which is that it decreases monotonically under incoherent operations (in resource theory of coherence, incoherent operations are those create only incoherent states for free, all coherent states become resource whose creations requires the costly implementation
of operations outside the set of incoherent operations, see \cite{baumgratz2014quantifying}).
  
We now state our desired result:
\begin{equation}
\Delta{W_f} = \Delta{E}+\beta^{-1} (\Delta{C_r}(\rho)-\Delta{S}(\Delta \rho)- \log(\frac{Z_t}{Z_0}))
\label{eq:18}
\end{equation}
where is easily derived from Eq.~(\ref{eq:13}) considering the definition of Eq.~(\ref{eq:17}).
The second term on the right hand site of Eq.~(\ref{eq:18}) involves respectively the difference between the relative entropy of coherence, the entropy of dephased state and the logarithm of partition function at the final and initial times of evolution that the latter arises from time variation of Hamiltonian and vanishes with time-independent Hamiltonians.
It is clear from this equation that the more coherent a quantum system is at a given time $t$ compared to its initial state, the more work can be extracted from it. For instance, this is the case when the system coherently exchanges energy with another quantum system or is coherently derived by an external field.
On the other hand, any incoherent operation decreases the quantum advantage on the work extraction by depleting the quantum resource of coherence. This case can be exemplified by a decoherence process that induces an open quantum system to lose its coherence to the environment.
In addition and in line with the work of Shi \emph{et al.} \cite{shi2022entanglement}, one can recognize the Eq.~(\ref{eq:18}) as a sum of two parts: a coherent part includes the change in the relative entropy of coherence $\Delta{W_{f}^{c}}=\beta^{-1} \Delta{C_r}(\rho)$ and an incoherent part includes other terms $\Delta{W_{f}^{i}}= \Delta{E} -\beta^{-1}(\Delta{S}(\Delta \rho)- \log(\frac{Z_t}{Z_0}))$. 
The charging power can also be decomposed into coherent and incoherent parts, $P(t)= P^{c}(t)+P^{i}(t)$ with
\begin{equation}
 P^{c}(t)=\beta^{-1} (\frac{\partial C_r(\rho(t))}{\partial t})
 \label{eq:19}
\end{equation}
and
\begin{equation}
 P^{i}(t)=\frac{\partial E} {\partial t} - \beta^{-1} (\frac{\partial S(\Delta{\rho(t))}}{\partial t}-\frac{\frac{\partial Z_t}{\partial t}}{Z_t}).  
 \label{eq:20}
\end{equation}
Eq.~(\ref{eq:19}) shows the revival of quantum coherence during the evolution, associated with $P^{c}(t) >0$, enhances the charging power and can have a dominant effect at higher temperatures.

\section{Examples}
In this section, we discuss two examples to illustrate the utility of the above results.
Examples are considered as different types of quantum batteries,
an interesting case study in the field of quantum thermodynamics to investigate the potential efficiency that may be acquired on energy storage of a battery by quantum mechanical phenomena. We employ Eqs.~(\ref{eq:16}) and (\ref{eq:19}) to demonstrate the thermodynamics advantage of quantum non-Markovianity in example 1 and quantum coherence in example 2 on the charging power evolution of the related quantum batteries.
\subsection{Example 1: Two-qubit battery-charger model}
In the first example, we consider a two-qubit system as a battery-charger model in which qubit 1 as a quantum battery is directly charged through interaction with qubit 2 as a charger while both qubits are simultaneously exposed to a common bosonic environment.
The Hamiltonian of the system is given by $\emph{H}=\emph{H}_0 + f(t)\emph{H}_I$, with
\begin{equation}
    \emph{H}_0 = \sum_{i=1}^{2}\omega_{0}\sigma_{i}^{+}\sigma_{i}^{-}+\sum_k\omega_{k} a_{k}^{\dagger}a_{k}
    \label{eq:21}
\end{equation}

and
\begin{equation}
    \emph{H}_I =(\alpha_1 \sigma_1^+ +\alpha_2 \sigma_2^+ ) \sum_k (g_{k}a_{k}) + h.c
    \label{eq:22}
\end{equation}
 where $\emph{H}_0$ includes free Hamiltonians of two qubits and environment with $\omega_0$ and $\sigma_{i}^{\pm}$ 
 being respectively for transition frequency and Pauli raising and lowering operators for i-th qubits. Also, $\omega_k$, $a_{k}^{\dagger}$ and $a_k$ are frequency, creation operator and annihilation operator of k-th mode of environment.
  $\emph{H}_I$ is the interaction Hamiltonian of two qubits with a common environment in which $g_k\alpha_i $ denotes 
the coupling constant between the i-th qubit and k-th mode of environment.
$f(t)$ is a dimensionless function that controls the coupling and decoupling of the charger to the battery with accepting respectively the value 1 
in charging time [0,$\tau$] and 0 elsewhere.
This model was solved analytically in one \cite{maniscalco2008protecting,francica2009off} and two \cite{mazzola2009sudden} excitation spaces and was first studied as a quantum battery-charger model in \cite{tabesh2020environment}. For the environment being initially in the vacuum state, we consider an initial state of the form
\begin{equation}
    |\psi(0)> = (c_{01} |1>_1 |0>_2 + c_{02} |0>_1 |1>_2) \otimes |0>_E
\end{equation}
where its time evolution can be expressed in terms of one excitation basis states as follows:
\begin{multline}
|\psi(t)> = c_1(t) |1>_1 |0>_2 |0>_E+ c_2(t) |0>_1 |1>_2 |0>_E\\ 
+ \sum_k c_k (t) |0>_1 |0>_2 |1>_E
\end{multline}
$|1>_E$ is the state of the environment with only one excitation
in the k-th mode.

 The reduced density matrix for the quantum battery can be thereby written as :
\begin{equation}
    \rho_1 (t) = |c_1(t)|^2 |1><1|_1 + (1-|c_1(t)|^2) |0><0|_1 
    \label{eq:25}.
\end{equation}
We assume in the following that the bath has a Lorentzian spectral density
\begin{equation}
    J(\omega) = \frac{W^2} {\pi} \frac{\lambda} {(\omega -\omega_0)^2+\lambda^2}
\end{equation}
where $\lambda$ is the width of the Lorentzian spectrum, which is connected to the environment correlation time $\frac{1} {\lambda}$ and W is an effective coupling strength related, in the limit of $\lambda\to0$, to Rabi frequency of two qubits $\Omega$. 
To distinguish the non-Markovian from Markovian dynamics we employ dimensionless parameter $\emph{R}=\frac{\Omega}{\lambda}$ that shows the Markovian and non-Markovian effects can be dominant respectively when $\emph{R}\ll 1$ and $\emph{R}\gg 1$ \cite{tabesh2020environment}.

We here suppose the initial state of the system as $ |\psi(0)> = |0>_1 |1>_2$ implying that the battery is empty and the charger has its maximum amount of energy before the charging process.
The details of the solution and analytical expression for amplitude $|c_1(t)|^2$ can be found in \cite{maniscalco2008protecting},
whereby the state of the battery is easily determined by Eq.~(\ref{eq:25}). The Gibbs state is reached by $|c_1(t)|^2=\frac{e^{-\beta \omega_{0}}}{Z}$ and $1-|c_1(t)|^2=\frac{1}{Z}$

\begin{figure*} [hbt!]
\includegraphics[width=7in, height=4in]{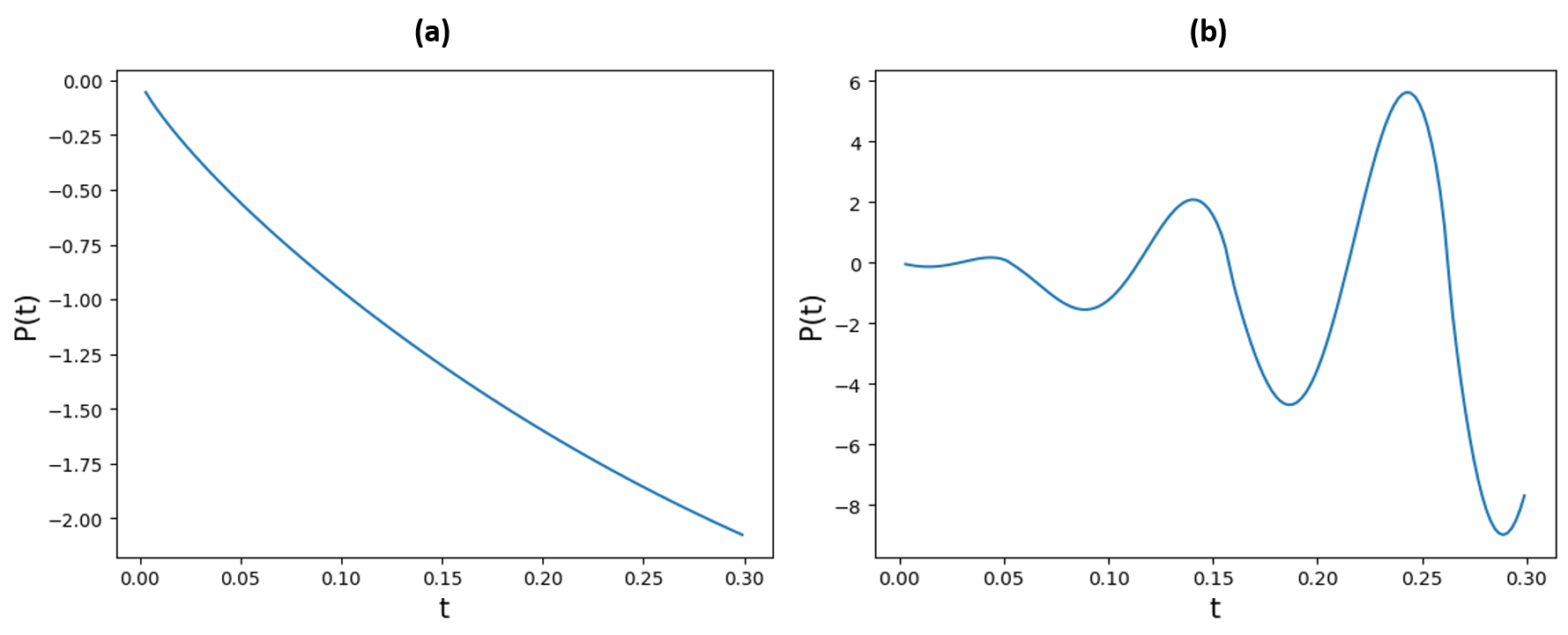}
\caption{\label{fig:1}Time evolution of charging power $P(t)$ of quantum battery in example 1 . (a) and (b) depict $\emph{R}=0.3$ and $\emph{R}=30$ respectively. The model parameters are set as $\omega_0 = 1$, $\beta =0.1$, $\lambda = 1$ and $\alpha = \frac{1}{\sqrt{2}}$.}
\end{figure*}

We then apply Eq.~(\ref{eq:16}) to calculate the charging power of the battery during the time evolution. The results have been plotted in Fig.~\ref{fig:1} for two different values of $\emph{R}$ taken as 0.3 and 30 to enable us to better compare the effects of Markovian and non-Markovian dynamics on the calculated results. With $\emph{R}=0.3$, as we would expect the dynamics to become mostly Markovian, Fig.~\ref{fig:1}(a) shows that the charging power decreases monotonically over the given time interval as a result of the monotonically decreasing non-Markovian characteristic of the dynamics (measured by $I$ in Eq.~(\ref{eq:15})). On the other hand, for $\emph{R}=30$, Fig.~\ref{fig:1}(b) shows revivals in the charging power of the battery in consequence of the increase in non-Markovianity ($I$) demonstrating that the system interpolates between Markovian and non-Markovian regimes. These results exhibit obviously that the charging power is enhanced by virtue of non-Markovian effects on the dynamics of the quantum battery a fact that is well expressed in Eq.~(\ref{eq:16}) by establishing a direct connection between two desired quantities $P$ and $I$. 
\subsection{Example 2: Two-qubit battery–one photon charger model}
To explain the role of quantum coherence, we consider in example 2 a two-qubit battery model in which the battery has been charged by coherence mediated by one photon mode. We investigate this model in two cases: in the first, the model is assumed as a close
system in which the battery and the photon evolve unitarily. In the second, we regard the detrimental effects of spontaneous emission brought on by the interaction of the battery with a Markovian environment. The Hamiltonian of the system is given by $\emph{H}=\emph{H}_0 + f(t)\emph{H}_I$, with
\begin{equation}
 \emph{H}_0 = \sum_{i=1}^{2}\omega_{0}\sigma_{i}^{+}\sigma_{i}^{-}+\omega_{p} a_{p}^{\dagger}a_{p}
\end{equation}
and
\begin{equation}
    \emph{H}_I =g (\sigma_1^+ + \sigma_2^+ )  (a_{p}) + h.c
\end{equation}
where p denotes the photon mode and all operators and parameters are defined before in Eqs.~(\ref{eq:21}) and (\ref{eq:22}). 
The dynamics of the quantum battery is described after tracing out the photon degree of freedom by the Schrödinger equation
\begin{equation}
 \frac{\partial \rho_b (t)} {\partial t} = i\ Tr_{p} ( [\rho,\emph{H}] )
 \label{eq:29}
\end{equation}
in the first case, and by a master equation \cite{ficek2002entangled,dzsotjan2010quantum}:
\begin{multline}
 \frac{\partial \rho_b (t)} {\partial t} = i\ Tr([\rho,\emph{H}^{'}])+\\ \sum_{i,j=1,1}^{2,2} \frac{\gamma_{ij}}{2} (2 \sigma_{i}^{-} \rho_b\ \sigma_{j}^{+}-\sigma_{i}^{+}\sigma_{j}^{-}\rho_b -\rho_b \sigma_{i}^{+}\sigma_{j}^{-})
 \label{eq:30}
\end{multline}
in the second case. Here, $\emph{H}^{'}$ is lamb shift Hamiltonian describing the effective interaction between two qubits:
\begin{equation}
 \emph{H}^{'} = \sum_{i=1}^{2}\omega_{0}\sigma_{i}^{+}\sigma_{i}^{-} + g_{12}(\sigma_{1}^{+}\sigma_{2}^{-}+\sigma_{2}^{+}\sigma_{1}^{-})
\end{equation}
with $g_{12}$ being the coupling constant. 
Also, $\gamma_{ii}$ is the spontaneous emission rate for i-th qubit while $\gamma_{ij}$ represents the contribution from two-qubit interactions. Here, we assume $\gamma_{11}=\gamma_{22}=\gamma$ and use $g_{12}= g^2/\Delta$ and $\gamma_{12}\approx0$ in accordance with realistic situations in cavity quantum electrodynamics (CQED) experiments \cite{gonzalez2011entanglement} where $\Delta$ denotes the detuning between frequencies of qubits and photonic cavity mode.

\begin{figure*} [hbt!]
    \includegraphics[width=7in, height=5in]{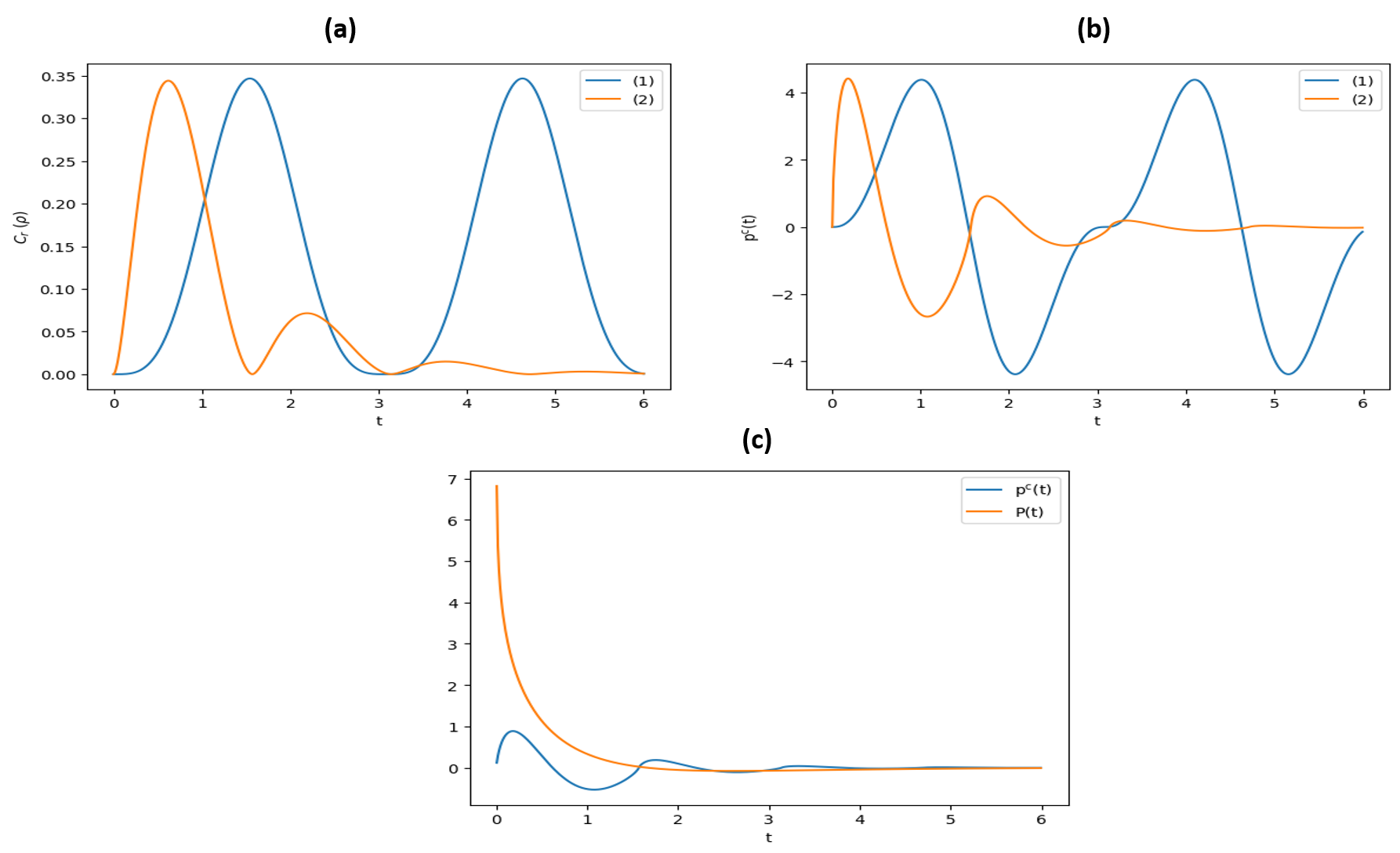}
    \caption{\label{fig:2}Time evolution of (a) relative entropy of coherence $C_{r}(\rho)$(b) coherent charging power $P^{c}(t)$ of quantum battery in example 2. (1) and (2) denote to cases 1 and 2 in the text.
Model parameters are taken as $g=1, \omega_0 =1, \omega_p = 2$ and $\beta = 0.1$. (c) depicts the coherent and total charging power of the quantum battery in case 2. Model parameters are the same as (b) and (c) except for $\beta = 1$ }
\end{figure*}

Once the dynamics of battery's state, $\rho_b (t) $, is obtained by numerically solving Eqs.~(\ref{eq:29}) and (\ref{eq:30}), the dynamics of its coherence is quantified by the quantum relative entropy of coherence, $C_{r}(\rho_b (t))$, defined by Eq.~(\ref{eq:17}).
We then employ Eqs.~(\ref{eq:19}) and (\ref{eq:20}) to calculate the coherent and the total charging power of the battery during the time evolution. 
The results for two cases are plotted in in Fig.~\ref{fig:2}.
In Fig.~\ref{fig:2}(a), one can see  
the battery in the first case exhibits a cycle of gain and loss of quantum coherence continuing indefinitely due to the unitarity of evolution.
In the second case, however, even though the battery reaches its maximum amount of quantum coherence more quickly than in the first case (thanks to the coherent part of Eq.~(\ref{eq:30}) in the first term), quantum coherence of battery vanishes as a result of decoherence process induced by interaction with Markovian bath.

The same behavior can be observed in Fig.~\ref{fig:2}(b) for the coherent part of the charging power of quantum battery as it indicates the oscillations with equal amplitude in case 1, while in case 2 the decoherence induced by the Markovian environment leads to null coherent charging power. Finally, in Fig.~\ref{fig:2}(c), it is shown that despite having initially markedly different values, the coherent and total charging powers nearly reach zero at the same time. This supports the main finding of Shi \emph{et al.} \cite{shi2022entanglement} that coherence must be attained during charging in order to extract work from a battery in its initial non-coherent state. The values of total charging power reveal that, notwithstanding early revivals in quantum coherence, the system loses its available work monotonically through interaction with the environment.

\section{Conclusions}
Examining how a system's quantum characteristics affect its thermodynamic performance is crucial to understanding quantum thermodynamics. This is particularly important in the context of work extraction from the quantum systems to design more efficient quantum thermodynamic devices like quantum heat engines and quantum batteries. The main objective of this study is to make one more contribution to addressing this problem by providing a formalism capable of highlighting the effective quantum parameters of the dynamics of extractable work of an open quantum system. 
Dynamics has been considered as a general quantum process which means the only limitation to dynamics is that it must be physically legitimate. Complete-positive and trace-preserving maps ensure this condition.
We have quantified the extractable work of the evolved system with the differences between its instantaneous non-equilibrium and equilibrium free energies that measure the maximum work that can be extracted from a quantum system when it is in contact with a thermal bath. We have established a fundamental thermodynamic equation that relates the change in internal energy of the quantum system to the change in its extractable work and entropy. We have demonstrated the salient fact about this equation appears when we write the entropy change of the system as a sum of two contributions stemming from reversible and irreversible processes. They respectively specify the heat flow and the change in extractable work of the system during the process.
We have demonstrated how this correspondence between irreversibility and extractable work can be used to explore the genuine quantum effects on the dynamics of the latter.  
In particular, this is accomplished by establishing a direct correspondence between the evolution of  extractable work and the evolution of the standard measures of quantum non-Markovianity and quantum coherence. 
Our results thus show that quantum non-Markovianity and quantum coherence are resourceful for the thermodynamic task of extracting work from an open quantum system evolving under a general quantum process and allow for a systematic study of the efficiency of these quantum phenomena for quantum thermodynamics applications.
We have illustrated our results by showing how the non-Markovian dynamics can improve the thermodynamic performance of an open quantum battery by increasing its charging power during the evolution. 
Both the battery and charger in this example are qubits. We have also demonstrated the same benefit is realized when we look at the role of quantum coherence induced by an optical charger in the charging power of a two-qubit battery.

\section*{Author Contributions}
\noindent Amin Mohammadi: Conceptualization, Writing – original draft. Afshin Shafiee: Supervision, Writing – review $\&$ editing

\section*{Conflict of Interest}
\noindent There are no conflicts of interest to declare.
\begin{acknowledgments}
\noindent We thank Dr. Ali Soltanmanesh for his constructive comments on the
presentation of the paper.
\end{acknowledgments}
\section*{REFERENCES}
\bibliography{REFERENCES}

\end{document}